\documentstyle[12pt,epsf,epsfig,wrapfig,latexsym,amsfonts,amssymb,amssymb,amsbsy,amsmath]{article}

\def\enquote#1{``#1''}
\begin{document}


\begin{center}
{\bfseries ELECTROMAGNETIC WAVE-PARTICLE\\ WITH SPIN AND MAGNETIC MOMENT}

\vskip 5mm
A.A. Chernitskii

\vskip 5mm
{\small
{\it A. Friedmann Laboratory for Theoretical Physics, St.-Petersburg, Russia;\\
State University of Engineering and Economics,\\ Marata str. 27, St.-Petersburg, Russia, 191002}\\
{\it
E-mail: AAChernitskii@mail.ru, AAChernitskii@engec.ru
}}
\end{center}

\vskip 5mm
\begin{abstract}
An axisymmetric static solution of a nonlinear electrodynamics is considered as a massive charged particle
with spin and magnetic moment. A linearization of the nonlinear electrodynamics around the static
solution is investigated. The appropriate problem for linear waves around the static solution is considered.
This wave part of the particle solution is considered to provide the appropriate wave properties for the particle.
It has been found that the right (experimentally proved) formula for frequency of this wave appears theoretically for the static solution with ring singularity.
\end{abstract}

\vskip 8mm \section{Introduction}

The field electromagnetic particle concept in the framework of a unified nonlinear electrodynamics is considered here.
In this approach a space-localized (soliton) solution of a nonlinear electrodynamics field model conforms to a physical
elementary particle. The term ``particle solution'' will be used here.
This theme was discussed in my articles
(see, for example, \cite{Chernitskii1999,Chernitskii2004a,Chernitskii2006a,Chernitskii2007a}).
The present work is concerned mainly with possible existence of a quick-oscillating part for a particle solution in
intrinsic coordinate system.
This quick-oscillating part in
intrinsic coordinate system of the particle will be a wave part for arbitrary coordinates.
 This wave part must provide the appropriate wave properties for the particle.

\section{Static electromagnetic particle with spin}

Axisymmetric static electromagnetic field configuration can have the spin defined as the full angular momentum of the electromagnetic field:
\begin{equation}
\label{63446234}
\mathtt{\mathbf{s}} = \int\pmb{\mathcal{M}}{\rm d}V
\;\;,
\end{equation}
where
$\pmb{\mathcal{M}}\doteqdot\mathbf{r}\times\pmb{\mathcal{P}}$ is an angular momentum density (spin density),
$\mathbf{r}$ is a position vector,
$\pmb{\mathcal{P}}\doteqdot\,(\mathbf{D}\times\mathbf{B})/4\pi$ is a momentum density (Poynting vector), $\mathbf{D}$ and $\mathbf{B}$ are electric and magnetic inductions.

The origin of the spin density is discussed in my articles. In particular, it was considered
three topologically different static field configuration with spin \cite{Chernitskii2007a}. These are configurations with two dyon, with singular disk,
and with singular ring.

Let us consider more closely the field configuration with singular ring.

\section{Toroidally symmetrical configuration\\ with singular ring}

A consideration of the static linear electrodynamics equations in toroidal coordinates $(\xi,\eta)$ gives the appropriate
solution with toroidal symmetry. This solution can include an electric and a magnetic parts. They can be represented with the help of toroidal harmonics
which are the spheroidal harmonics with half-integer index: $P_{n-\frac{1}{2}}^l(\cosh\xi)$, where $n$ and $l$ are integer.
To obtain the right behaviour of the electromagnetic field at infinity for a charged particle with
magnetic moment we must take the toroidal harmonics $P_{-\frac{1}{2}}^0(\cosh\xi)$, $P_{-\frac{3}{2}}^0(\cosh\xi)$ for
the electric field and $P_{-\frac{1}{2}}^1(\cosh\xi)$, $P_{-\frac{3}{2}}^1(\cosh\xi)$ for
the magnetic one. Because we intend to consider this solution as an initial approximation to a solution of a nonlinear
electrodynamics model, it is reasonable to take the condition of vanishing of two electromagnetic invariants near
the singular ring. This condition will be satisfied when the ratio between the electric and magnetic vector magnitudes tends to unit
near the ring.

We have the following appropriate solution of the linear electrodynamics:
\def\ellintE{\textstyle \rm \int\!\!\!\! {\scriptstyle E}}
\def\ellintK{\textstyle \rm \int\!\!\!\! {\scriptstyle K}}
\begin{equation}
\label{skeuhgf}
  \begin{array}{rcl}
    D_\xi &=&-\displaystyle{\frac{e}{\sqrt{2}\,\pi\,\rho^2}} \,\sqrt{\cosh\xi-\cos\eta}\,
    \,{\rm csch}\xi
    \left[2\left(\cosh\xi - \cos\eta\right) \ellintE
         \Bigl({\displaystyle -\sinh^2\frac{\xi}{2}} \Bigr)
         \right.\\
    && \quad\quad\quad\quad\quad\quad\quad \left.
    -\left(1- \cos\eta\right)\left(1+\cosh\xi\right)\ellintK\Bigl({\displaystyle -\sinh^2\frac{\xi}{2}}\Bigr)\right]
\quad, \\[15pt]
  D_\eta &=& -{\displaystyle \frac{e}{\sqrt{2}\,\pi\,\rho^2}} \,
\sqrt{\cosh\xi-\cos\eta}\;\ellintK\Bigl({\displaystyle -\sinh^2\frac{\xi}{2}}\Bigr)\,\sin\eta
    \quad,\\[15pt]
    H_\xi &=& -{\displaystyle \frac{e}{\sqrt{2}\,\rho^2}} \,i\,\sqrt{\cosh\xi-\cos\eta}\,P_{-\frac{1}{2}}^1(\cosh\xi)\,\sin\eta
\quad, \\[15pt]
    H_\eta & =&\displaystyle{\frac{e}{\sqrt{2}\,\rho^2}} \,i\,\sqrt{\cosh\xi-\cos\eta}\,
    \,{\rm csch}\xi\left[\left(\cosh\xi - \cos\eta\right) P_{-\frac{3}{2}}^1(\cosh\xi)\right.\\
    && \quad\quad\quad\quad\quad\quad\quad\quad\quad\quad\quad +\left.
    \left( \cos\eta\,\cosh\xi- 1 \right)P_{-\frac{1}{2}}^1(\cosh\xi)\right]\quad,
  \end{array}
\end{equation}
where $D_\xi$, $D_\eta$, $H_\xi$, $H_\eta$ are the physical components of the electric induction and magnetic strength vectors in toroidal coordinates, $\ellintK(m)$ and $\ellintE(m)$ are complete elliptic integrals of the first and second kinds accordingly (can be expressed by means of
$P_{-\frac{1}{2}}^0(\cosh\xi)$ and $P_{-\frac{3}{2}}^0(\cosh\xi)$),
$P_{-\frac{1}{2}}^l(z)$ and $P_{-\frac{3}{2}}^l(z)$ are associated Legendre functions with half-integer negative index.
Here $\rho$ is the radius of the ring and $e$ is a constant.

The field configuration (\ref{skeuhgf}) has the electric charge $e$ and the magnetic moment
\begin{equation}\label{123w11}
    \mu = \frac{e\,\rho}{2}\quad.
\end{equation}
About the definition of the electric charge and the magnetic moment see my article \cite{Chernitskii2006a}.

\section{Wave part of the particle solution}

In general case particle solution can include a time-periodic part.
Having in view the wave properties of the physical particles this time-periodic part must be considered as quick-oscillating one.
For the intrinsic coordinate system of the particle this quick-oscillating part can be a standing wave or a progressive wave around
a closed singular line (ring, for example).
A hyperbolic rotation of the particle solution with the quick-oscillating part gives the solution for moving particle having
evident wave properties.

The appropriate time-periodic solutions well known for the linear electrodynamics. In particular, they exist for considered
topologically different configurations of singularities: bidyon, disk, and ring.

For the nonlinear case we can investigate the time-periodic part by means of the linearization of the problem
around the appropriate static field configuration (which may not be exact solution).
The general method for the linearization is based on Frechet derivative for the nonlinear operator:
\begin{equation}
\nonumber
    \begin{array}{c}
    \mathcal{N}\mathbf{Y}=0\\
        \text{Nonlinear}\\
        \text{system of equations}
    \end{array}
    \quad
    \begin{array}{c}
      \to\\ \\
      \mathbf{Y}=\mathbf{Y}_0+\tilde{\mathbf{Y}}
    \end{array}\quad
    \begin{array}{c}
    \mathcal{N^\prime}(\mathbf{Y}_0)\tilde{\mathbf{Y}}=-\mathcal{N}\mathbf{Y}_0\\
        \text{Linearized around $\mathbf{Y}_0$}\\
        \text{system of equations}
    \end{array}
    \quad.
\end{equation}

This problem for point singularity was considered in my article \cite{Chernitskii2006b}.

\section{About circular frequency\\ of the time-periodic part of the particle solution}

Let us consider that the circular frequency of the time-periodic part of the particle solution is given by the known formula:
$m  = \hbar \,\omega \;\; (c = 1)$.

The magnetic moment of the Dirac spin one-half particle is $\mu   = \dfrac{{e\,\hbar }}{{2\,m }}$.

Thus we have the important formula for the circular frequency:
\begin{equation}\label{ddfvvfr}
\left. \begin{array}{l}
 \mu   = \dfrac{{e\,\hbar }}{{2\,m }}\,\,\, \Rightarrow \,\,\,\dfrac{{m }}{\hbar } = \dfrac{e}{{2\,\mu }}  \\[4ex]
 m  = \hbar \,\omega \,\,\, \Rightarrow \,\,\,\dfrac{{m }}{\hbar } = \omega  \\
 \end{array} \right\}\,\,\, \Rightarrow \,\,\,\omega  = \dfrac{e}{{2\,\mu}}
\end{equation}
Here the initial formulas have the experimental confirmations. Thus the obtained formula must be considered as experimental one.

\section{The advance for particle solution\\ with ring singularity}

The ring configuration gives birth to the appropriate periodic space boundary condition (on the ring).
Thus it would appear reasonable that we will have the time-periodic part of the particle solution in the form of wave
propagating along the ring.
The wave-length of the appropriate fundamental mode is
\begin{equation}\label{dwo93}
    \lambda = 2\,\pi\,\rho\quad.
\end{equation}

In general case the static part of the solution can modify the conditions of propagation for the wave part of the solution
(see my article \cite{Chernitskii1998b}). But for the special case of vanishing of the electromagnetic invariants for the static
part of the solution we can have the changeless speed of light $c=1$ for the wave part.

In this case the fundamental circular frequency for the time-periodic part is
\begin{equation}\label{sdflei}
    \omega=\dfrac{2\pi}{\lambda} = \dfrac{1}{\rho}\quad.
\end{equation}

Combining the formulas (\ref{123w11}) and (\ref{sdflei}) we have:
\begin{equation}\label{euiopop}
\left. \begin{array}{l}
 \mu=\dfrac{e\,\rho}{2}  \\[2ex]
 \omega  = \dfrac{1}{{\rho }} \\
 \end{array} \right\}\,\,\,\, \Rightarrow \,\,\,\omega  = \frac{e}{{2\mu }}\quad.
\end{equation}

As we see this formula coincides with the experimental formula (\ref{ddfvvfr}).

It should be noted that the appropriate field configuration with disk-shaped singularity (see, for example, \cite{Burinskii1974e}) 
has the magnetic moment described by formula
    $\mu=e\,\rho$
that is not leading in this approach to the right formula (\ref{ddfvvfr}).

\section{Conclusions}
Thus we must call our attention to the particle solutions with ring singularity.
Such particle solutions may represents the real physical elementary particles.

\providecommand{\href}[2]{#2}


\begin{thebibliography}{6}
\bibitem{Chernitskii1999}
A.A. Chernitskii, \enquote{Dyons and interactions in nonlinear (Born-Infeld) electrodynamics,} \emph{J. High Energy Phys.} \textbf{1999}, No 12, Paper 10, 1--34 (1999); [\href{http://arxiv.org/abs/hep-th/9911093}{{\tt hep-th/9911093}}].

\bibitem{Chernitskii2004a}
A.A. Chernitskii, \enquote{Born-Infeld equations,} in
  \emph{Encyclopedia of Nonlinear Science}, edited by
  A.~Scott, Routledge, New York and London, 2004, pp. 67--69;
 [\href{http://arxiv.org/abs/hep-th/0509087}{{\tt hep-th/0509087}}].


\bibitem{Chernitskii2006a}
A.A. Chernitskii, \enquote{Mass, spin, charge, and magnetic moment for
  electromagnetic particle,} in \emph{XI Advanced Research Workshop on High
  Energy Spin Physics (DUBNA-SPIN-05) Proceedings},
  edited by A.~V. Efremov, and S.~V. Goloskokov, JINR, Dubna,
  2006, pp. 234--239; [\href{http://arxiv.org/abs/hep-th/0509087}{{\tt hep-th/0603040}}].

 \bibitem{Chernitskii2007a}
A.A. Chernitskii, \enquote{The field nature of spin for electromagnetic particle,}
in \emph{Proceedings of the 17-th International Spin Physics Symposium (Kyoto, Japan, 2--7 Oktober 2006)},
  edited by K.~Imai, T.~Murakami, N.~Saito, and K.~Tanida, AIP Conference Proceedings {\bf 915},
  Melville, New York, 2007, pp. 264--267; [\href{http://arxiv.org/abs/hep-th/0509087}{{\tt hep-th/0611342}}].


\bibitem{Chernitskii2006b}
A.A. Chernitskii, \enquote{Linear waves around static dyon solution of nonlinear (Born-Infeld)   electrodynamics,}
 \emph{Hadronic Journal}, \textbf{29}, 497--528  (2006); [\href{http://arxiv.org/abs/hep-th/0509087}{{\tt hep-th/0602079}}].

\bibitem{Chernitskii1998b}
A.A. Chernitskii, \enquote{Light beams distortion in nonlinear electrodynamics,} \emph{J. High Energy Phys.} \textbf{1998}, No 11, Paper 15, 1--5
  (1998); [\href{http://arxiv.org/abs/hep-th/0509087}{{\tt hep-th/9809175}}].

\bibitem{Burinskii1974e}
A.~Burinskii, \enquote{Microgeon with spin,}
\emph{Sov. Phys. JETP} \textbf{39}, 193 (1974).

\end{thebibliography}
\end{document}